\documentclass[intlimits,twoside,a4paper]{article}

\usepackage{amsmath,amssymb}
\usepackage{graphicx}

\usepackage[cp1251]{inputenc}

\usepackage[eqsecnum]{cmpj3}

\issue{2017}{20}{4}{43708}
\doinumber{10.5488/CMP.20.43708}
\title[Investigation on TiO$_2$ nanostructures]%
{Sensing behavior of acetone vapors on TiO$_2$ nanostructures --- application of density functional theory }%
\author[V. Nagarajan, S. Sriram, R. Chandiramouli]{V. Nagarajan, S. Sriram, R. Chandiramouli\thanks{Corresponding author}}
\address{
School of Electrical and Electronics Engineering, Shanmugha Arts Science Technology and Research Academy (SASTRA) University, Tirumalaisamudram, Thanjavur, Tamil nadu --- 613 401, India 
}

\date{Received July 13, 2017, in final form September 5, 2017}
\begin{document}

\maketitle

\begin{abstract}
The electronic properties of TiO$_2$ nanostructure are explored using density functional theory. The adsorption properties of acetone on TiO$_2$ nanostructure are studied in terms of adsorption energy, average energy gap variation and Mulliken charge transfer. The density of states spectrum and the band structure clearly reveals  the adsorption of acetone on TiO$_2$ nanostructures. The variation in the energy gap and changes in the density of charge are observed upon adsorption of acetone on n-type TiO$_2$ base material. The results of DOS spectrum reveal that the transfer of electrons takes place between acetone vapor and TiO$_2$ base material.  The findings show that the adsorption property of acetone is more favorable on TiO$_2$ nanostructure. Suitable adsorption sites of acetone on TiO$_2$ nanostructure are identified at atomistic level. From the results, it is confirmed that TiO$_2$ nanostructure can be efficiently utilized as a sensing element for the detection of acetone vapor in a mixed environment.
\pacs 71.15.Mb 
\end{abstract}

\section{Introduction}

The expansion of industries in recent years leads to emission of hazardous gases and vapors into the atmosphere. Moreover, volatile organic compounds (VOCs) are a major source of environmental pollutants and cause a serious impact on humans. For instance, acetone (CH$_3$COCH$_3$) is a chemical reagent utilized in laboratories and industries. Besides, this compound is widely used in purifying paraffin, dissolving plastics and in pharmaceutics. Acetone may cause damages to human noses, eyes and central nervous system when the permissible exposure limit exceeds 1000 parts per million (ppm) according to Occupational Safety and Health Administration \cite{1}. The high exposure to acetone to humans may cause mood swings, respiratory irritation and nausea. In addition, breathing acetone in high ppm value may cause dizziness, respiratory tract irritation and loss of strength \cite{2}. Furthermore, acetone is also highly inflammable. Meanwhile, acetone was found to be the final product for added ketone bodies’ metabolism \cite{3}. Among the transition metal oxide semiconductor, titanium dioxide (TiO$_2$) is extensively investigated as a key material for technological application and fundamental research in the semiconductors, solar cell \cite{4} and lithium-ion batteries \cite{5} owing to its excellent chemical stability and low cost \cite{6}. TiO$_2$ is also a promising candidate in the field of gas-sensor, photovoltaic, energy storage and photocatalysis due to its photocatalytic properties, long-term stability and low toxicity \cite{7,8}. Besides, TiO$_2$ polymorphs are mainly classified into three types, namely anatase, rutile and brookite with corresponding space groups namely I4$_1$/amd-$D_{4h}^{19}$ (tetragonal), P4$_2$/mnm-$D_{4h}^{14}$ (tetragonal) and pbca-$D_{2h}^{15}$ (orthorhombic). Besides, only the first two crystal systems play a vital role in industrial applications. Experimental work on brookite crystal system is constrained owing to its difficulty in preparation \cite{9}. The optical energy gap values of anatase, rutile and brookite TiO$_2$ crystal structure are reported as 3.4~eV \cite{10}, 3.0~eV \cite{11} and 3.3~eV \cite{12}, respectively. The density functional theory (DFT) method has been widely used by the researchers to study rutile and anatase phases of TiO$_2$ nanostructure \cite{13}. The previous reports on TiO$_2$ electronic properties were presented in the literature using hybrid-functional schemes \cite{14}. 
Bhowmik et al. \cite{15} have proposed TiO$_2$ nanotubes as a good acetone sensor and the authors observed that the response of the sensor reaches $3.35\% $ against 1000~ppm acetone. Chen et al. \cite{16} have reported the sensing performance of acetone based on nanoporous morphology of TiO$_2$ using a facile hydrothermal method. Rella et al. \cite{17} reported about the sensing properties of ethanol and acetone vapors on TiO$_2$ nanoparticles prepared from pulsed laser deposition. Chen et al. \cite{18} studied the adsorption properties of acetone on pristine and transition metal doped TiO$_2$ clusters using DFT study. Thus, from the previous literatures, to our knowledge, there are only limited reports based on DFT method to study the adsorption properties of acetone on rutile TiO$_2$ nanostructure. The motivation behind the present work is to investigate acetone adsorption properties on TiO$_2$ nanostructure and to identify the most suitable adsorption site at the atomistic level.

\section {Computational methods}
The electronic and adsorption properties of acetone on TiO$_2$ nanostructures are studied using DFT method utilizing SIESTA package \cite{19}. The atomic position in TiO$_2$ nanostructures has been optimized to their ground state by decreasing their Hellman-Feynman forces \cite{20,21} supported with conjugated gradient algorithm. Moreover, all interatomic forces of rutile TiO$_2$ nanostructures are observed to be less than 0.05~eV/{\AA} \cite{22}. The generalized gradient approximation (GGA) combined with Perdew-Burke-Ernzerhof (PBE) exchange correlation functional is utilized to investigate the electron-electron interaction \cite{23,24}. The energy cut-off of plane-wave basis set was adjusted to 500~eV with energy convergence of $10^{-5}$~eV. The atomic positions in TiO$_2$ nanostructure were relaxed until the force of 0.02~eV/{\AA} is achieved. The $k$-points in Brillouin zones under the Monkhorst-pack scheme \cite{25}, are kept as $10 \times 10 \times 10$ $k$-points with $3 \times 3 \times 3$ super-cell size in the present study. The density of states, charge density, band structure and electron localization function (ELF) of rutile TiO$_2$ nanostructure were calculated with the help of SIESTA code with Monkhorst-Pack $k$-point meshes of $10^{-3}$~{\AA}$^{-1}$. The acetone adsorption properties on rutile TiO$_2$ nanostructure is also studied using SIESTA package. Furthermore, the electronic wave function of Ti and O atoms is expressed by basis set, which is directly related with the numerical orbitals. The double zeta polarization (DZP) basis set is utilized for relaxation of rutile TiO$_2$ nanostructure in the present work \cite{26}.

\section {Results and discussion}
\subsection {Electronic properties of rutile TiO$_2$ nanostructure}

\begin{figure}[!t]
\centerline{\includegraphics[width=0.7\textwidth]{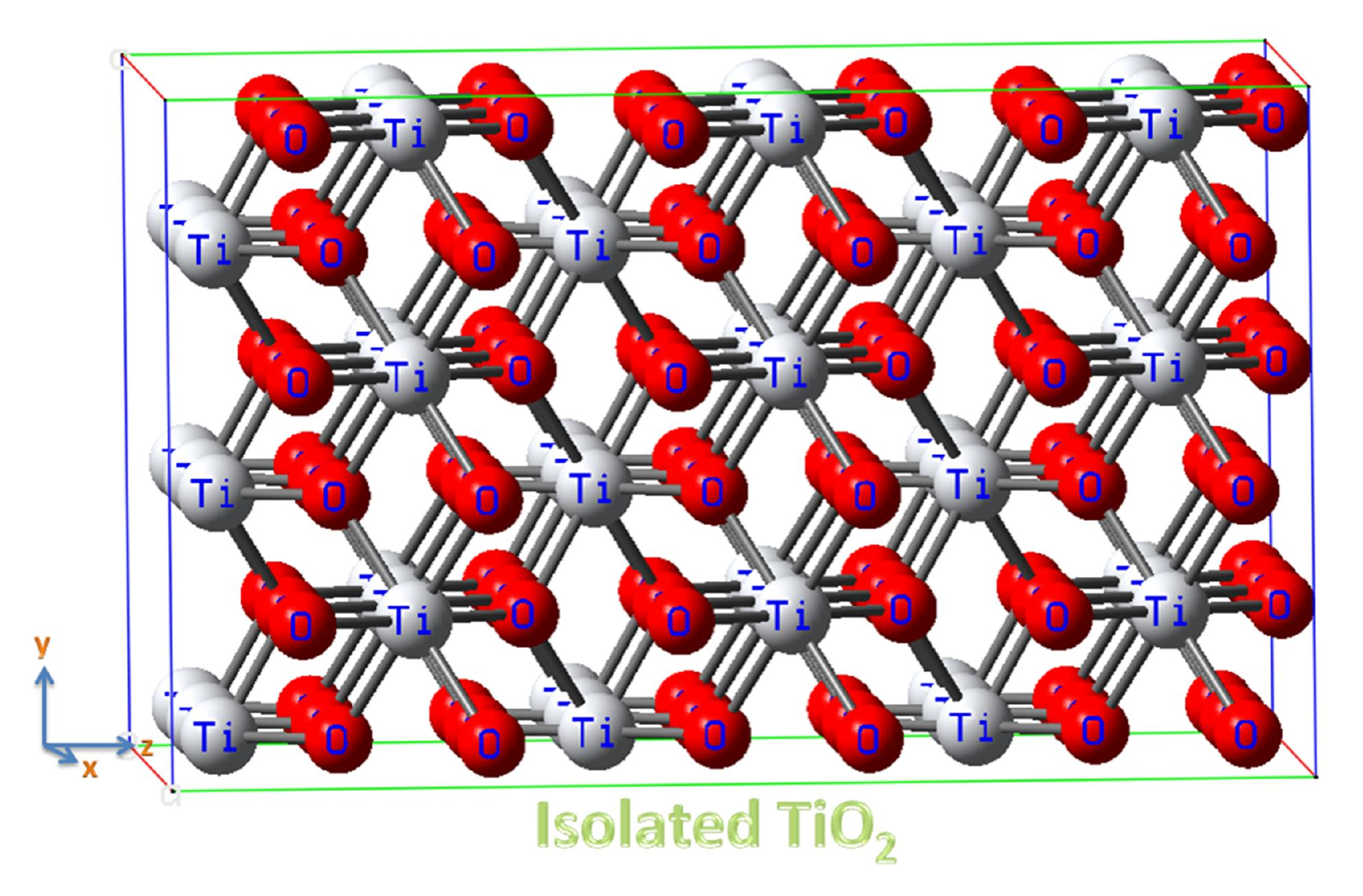}}
\caption{(Color online) Schematic diagram of an isolated TiO$_2$ nanostructure with periodic boundary condition.} \label{fig-s1}
\end{figure}

The present work concentrates on the investigation of electronic properties and adsorption behavior of acetone on TiO$_2$ nanostructure. Figure~\ref{fig-s1} represents the schematic diagram of TiO$_2$ nanostructure with the periodic boundary condition (PBC). The optimized lattice constant and Wyckoff atomic positions, including coordinates of TiO$_2$ nanostructure are tabulated in table~\ref{tab1}. In this work, we have chosen a tetragonal --- rutile TiO$_2$ nanostructure (P4$_2$/mnm) for calculating the electronic properties and adsorption behavior of acetone on TiO$_2$ nanostructure.

\begin{table}[!t]
\caption{\label{tab1} The Wyckoff atomic positions of TiO$_2$ nanostructure.}
\vspace{2ex}
\centering
\begin{tabular}{|c|c|c|l|}
\hline\hline
Structure& Space group & Lattice& Wyckoff atomic positions \\
\hline\hline
Tetragonal --- rutile TiO$_2$& P4$_2$/mnm (136)&$a = 4.59$~{\AA}   &Ti: $2a$ (0, 0, 0) \\

&  &$b = 4.59$~{\AA} &Ti: $16k$ (0.50, 0.50, 0.50) \\

&  & $c = 2.96$~{\AA} & O: $8i$ (0.31, 0.31, 0) \\

&& & O: $8i$ (0.70, 0.70, 0) \\

&& & O: $16k$ (0.20, 0.81, 0.50) \\

&& & O: $16k$ (0.81, 0.20, 0.50) \\
\hline\hline
\end{tabular}
\end{table}

\begin{figure}[!t]
\centerline{\includegraphics[width=0.7\textwidth]{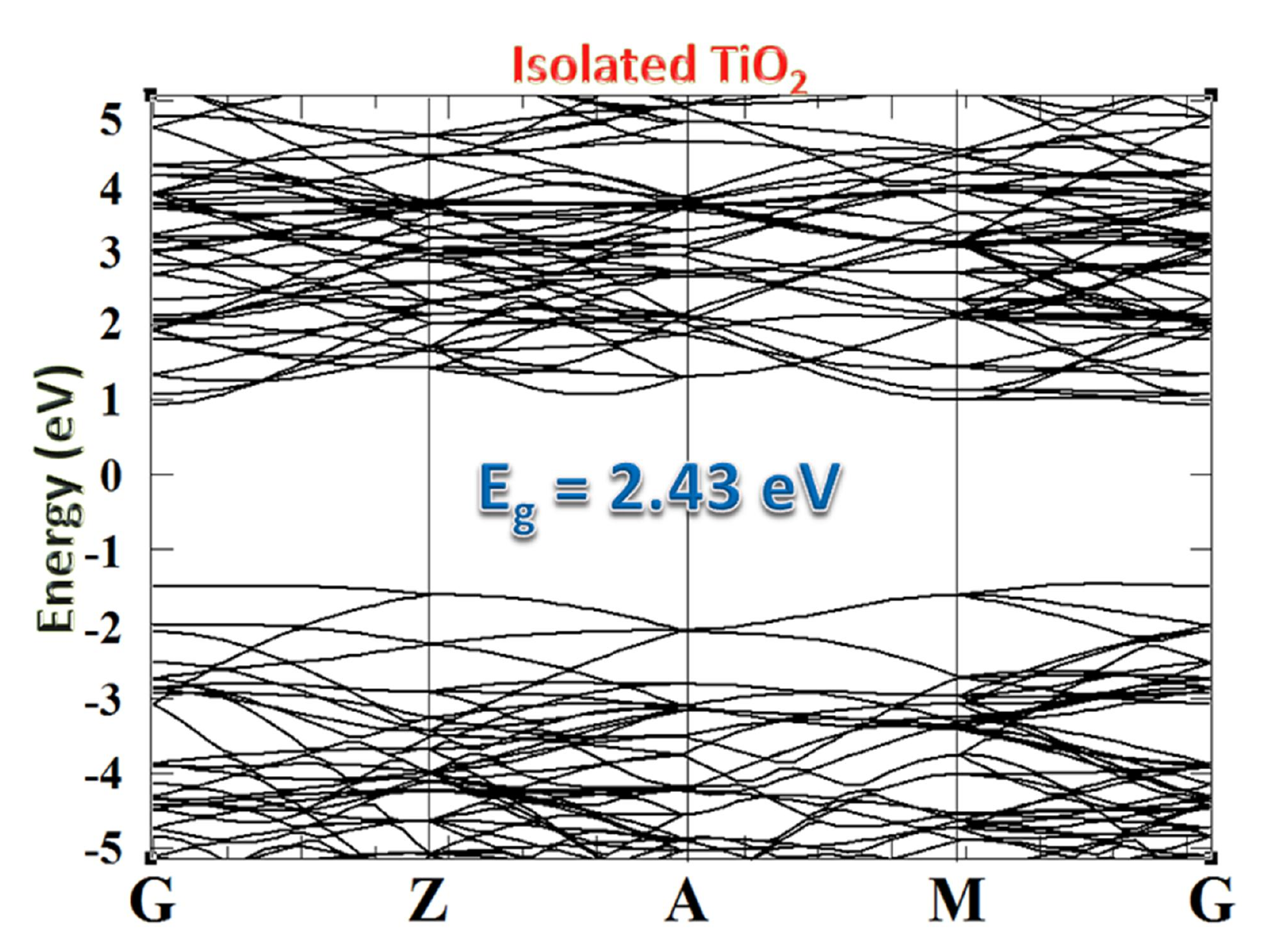}}
\caption{(Color online) Band structure of an isolated TiO$_2$ nanostructure.} \label{fig-s2}
\end{figure}

The electronic properties of TiO$_2$ base material are described in terms of band structure \cite{27}. Figure~\ref{fig-s2} illustrates the band structure of an isolated TiO$_2$ nanostructure. The band gap of TiO$_2$ base material is observed along the gamma point (G) and it is observed that the band gap value of an isolated TiO$_2$ nanostructure is found to be 2.43~eV with indirect gap. Moreover, the channels along the conduction band minimum and the valence band maximum are out of phase along the gamma point, which infers the indirect band gap. Furthermore, the calculated band gap of TiO$_2$ nanostructure is in good agreement with the previously reported work \cite{11}. In addition, based on the selection of exchange-correlation (XC) functional, the electronic properties of TiO$_2$ nanostructure can be fine-tuned. Previously, Deak et al. \cite{14}  reported the electronic band gap of TiO$_2$ with hybrid HSE06 functional for both anatase and rutile crystal structure, which are overestimated from the present work. Table~\ref{table 2} shows that the energy band gap value of an isolated TiO$_2$ nanostructure with different exchange correlation functional such as PBEsol, BLYP, RPBE, PBE and revPBE \cite{28}. Kaloni and co-workers \cite{29,30,31,32,33} have suggested that the electronic and  structural properties of 2D nanosheets and organic compounds are in oligomer form, which can be fine-tuned with the interaction of transition metal atoms.

\begin{table}[!t]
\caption{ Energy gap calculation for isolated TiO$_2$ nanostructure with various exchange-correlation functional.}
\label{table 2}
\vspace{2ex}
\centering
\begin{tabular}{|c|c|c|}
\hline\hline
XC functional&$E_{\text g}$ (eV) &Type\\
\hline\hline
PBE &2.43 &GGA\\
\hline
BLYP &2.50 &GGA \\
\hline
PBEsol &2.38 &GGA \\
\hline
RPBE &2.56 &GGA \\
\hline
revPBE &2.48 &GGA \\
\hline\hline
\end {tabular}
\end{table}

\begin{figure}[!t]
\centerline{\includegraphics[width=0.99\textwidth]{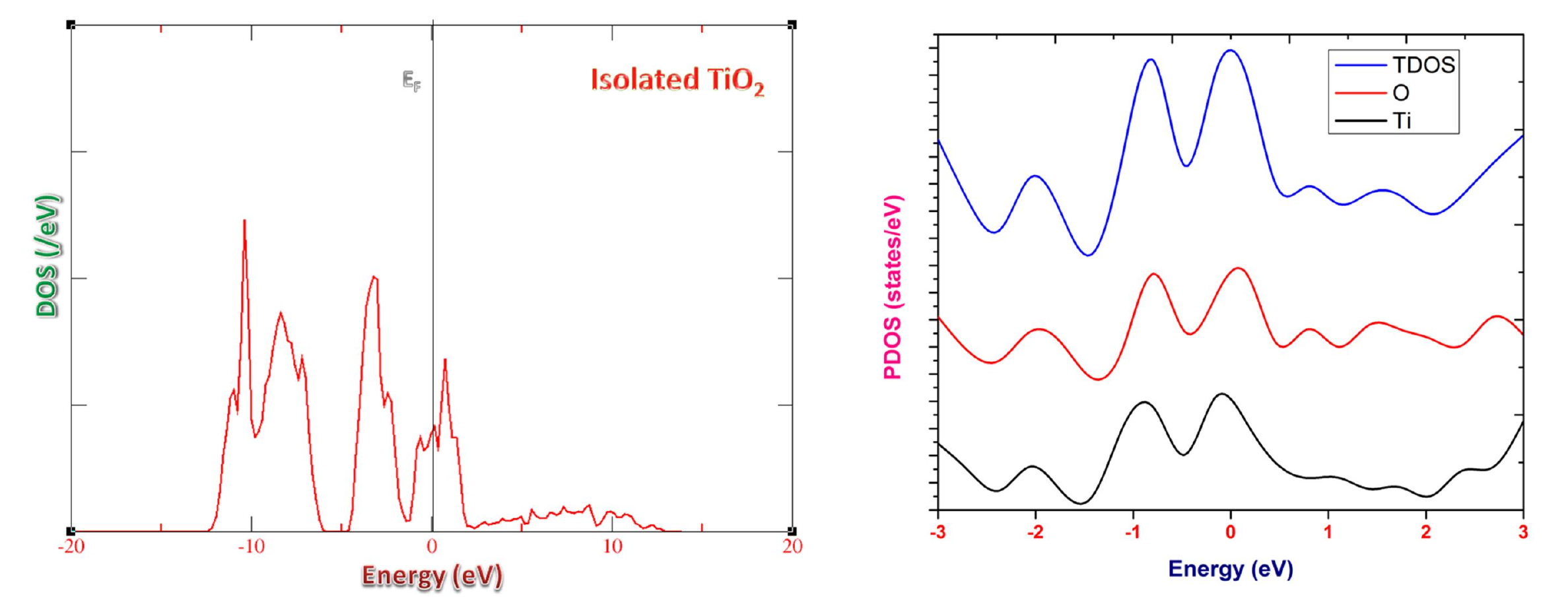}}
\caption{(Color online) Density of states (DOS) spectrum of an isolated TiO$_2$ nanostructure.} \label{fig-s3}
\end{figure}

The band structure of TiO$_2$ nanostructure is underestimated in the present work since the density functional theory method with GGA/PBE exchange correlation functional is utilized to calculate the electron-electron interaction in their ground state. Moreover, the electronic and adsorption properties of acetone in TiO$_2$ nanostructure cannot be disturbed due to underestimation of the band gap, since the isolated TiO$_2$ base material is compared with acetone adsorbed TiO$_2$ nanostructures. The density of states (DOS) spectrum gives the insights on localization of charges in different energy intervals along TiO$_2$ nanostructure. The visualization of band structure, PDOS and DOS spectrum are shown in figures~\ref{fig-s2} and \ref{fig-s3}, respectively. 

The peak maxima are recorded near the Fermi energy level (EF), which is one of the favorable conditions for the adsorption of target vapor/gas molecules. Thereby, the free electrons can easily transfer between TiO$_2$ base material and acetone molecules. The peak maximum in different energy intervals arose owing to the orbital overlapping between Ti and O atoms in TiO$_2$ nanostructure. Generally, for chemiresistive type of vapor/gas sensors, metal oxide semiconducting materials are preferred owing to the fact that the transfer of electrons is facilitated between target gas/vapor and TiO$_2$ nanostructure  and the changes in the resistance can also be observed. From the observations of band structure and DOS spectrum of TiO$_2$ nanostructure it is inferred that TiO$_2$ material can be used as a base material for the possible application of chemical nanosensors.
\newpage

\subsection {Adsorption properties of acetone on TiO$_2$ nanostructure}

\begin{figure}[!b]
\vspace{-3mm}
\centerline{\includegraphics[width=0.99\textwidth]{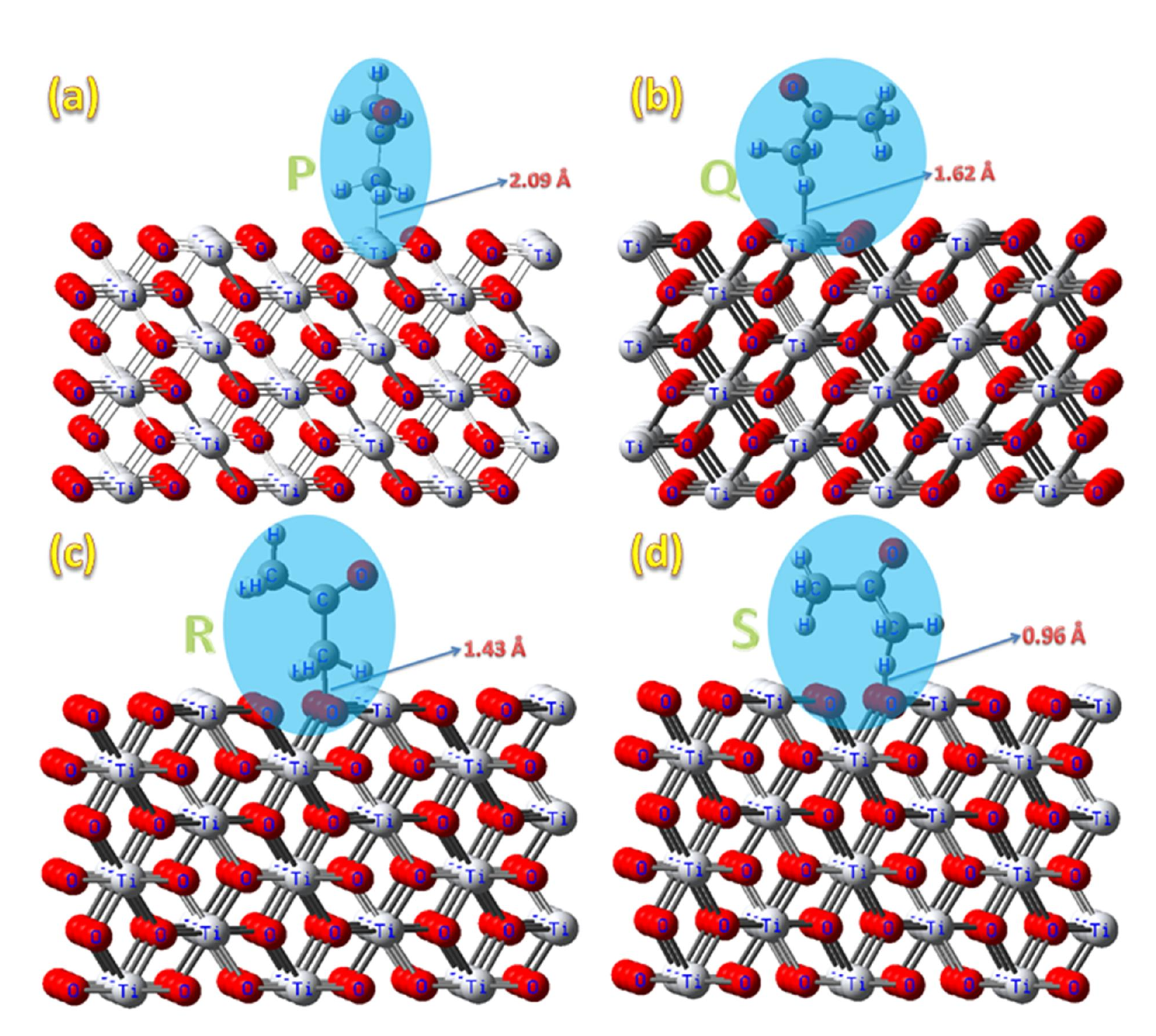}}
\caption{(Color online) (a)--(d) Adsorption of acetone on position P, Q, R and S.} \label{fig-s4}
\end{figure}

In the initial stage of acetone adsorption study on TiO$_2$ base material, acetone vapor should be studied in vapor phase. The bond length between Ti and O atom in TiO$_2$ nanostructure is 1.98~{\AA}. Figure~\ref{fig-s4}~(a) refers the adsorption of carbon atom in acetone molecules adsorbed on Ti atom in TiO$_2$ nanostructure and it is referred to as position P. Figure~\ref{fig-s4}~(b) illustrates the adsorption of H atom in acetone molecules adsorbed on Ti atom in TiO$_2$ nanostructure and it is referred to as position Q. Similarly, positions R and S refer to the adsorption of C and H atom in acetone molecules adsorbed on O atom in TiO$_2$ nanostructure as shown in figure~\ref{fig-s4}~(c) and (d), respectively. 
The adsorption energy of acetone on TiO$_2$ nanostructure can be calculated using equation~(\ref{eq1})

\begin{equation}
\label{eq1}
E_{\text{ad}} = [E(\text{TiO}_2/\text{CH}_3\text{COCH}_3) - E(\text{TiO}_2) - E(\text{CH}_3\text{COCH}_3) + E (\text{BSSE})],
\end{equation}
where $E(\text{TiO}_2/\text{CH}_3\text{COCH}_3)$ refers to the energy of TiO$_2$/CH$_3$COCH$_3$ complex. $E$(CH$_3$COCH$_3$) and $E$(TiO$_2$) refers to the isolated energy of CH$_3$COCH$_3$ and  TiO$_2$ molecules, respectively. BSSE represents the basis set superposition error using counterpoise techniques in order to eliminate the overlapping effect on basis functions during calculations.

When acetone molecules get adsorbed on TiO$_2$ base material, negative values of adsorption energy ($E_{\text{ad}}$) indicate a strong adsorption of acetone on TiO$_2$ nanostructure. Besides, in the present work, for all positions namely P--S exhibits a negative value of $E_{\text{ad}}$. This clearly confirms the adsorption of acetone molecules on TiO$_2$ nanostructure. The adsorption energy of TiO$_2$ nanostructure for position P--S is observed to be $-1.48$~eV, $-1.46$~eV, $-1.46$~eV and $-1.29$~eV, respectively. In addition, the band gap of TiO$_2$ nanostructure gets decreased due to the adsorption of acetone molecules on TiO$_2$ base material owing to the interaction of target VOCs with base material. Thus, the conductivity of TiO$_2$ nanostructure increases. The resistance of TiO$_2$ base material decreases when it is exposed to the reducing vapors such as acetone.  The trend in the changes of resistance upon exposure towards acetone vapor is in good agreement with the reported work of Sun et al. \cite{34}. The changes in band gap of TiO$_2$ nanostructure for positions P--S are found to be 2.32~eV, 1.32~eV, 0.82~eV and 0.68~eV, respectively. Besides, the variation in the energy gap and adsorption energy supports that TiO$_2$ nanostructure can be used for the detection of acetone vapor. Rella et al. \cite{17} reported about the detection of ethanol and acetone vapors using TiO$_2$ nanoparticles synthesized by pulsed laser deposition method.  Chen et al. \cite{16} studied the acetone sensing performance of nanoporous titanium dioxide using a facile hydrothermal method. Epifani et al. \cite{35} reported about the design of an acetone sensor based on titanium dioxide nanocrystals functionalized with tungsten oxide species. From the literature, it is evident that TiO$_2$ nanomaterial can be utilized as a two probe device for the detection of VOCs upon adsorption, which gives rise to change in the current. The variation in the current is directly proportional to the concentration of acetone molecules present in the atmosphere. From the previous reports, it is inferred that TiO$_2$ nanostructure can be efficiently used as acetone vapor sensor, which also strengthens the present work. Besides, the present investigation strongly confirms the adsorption of acetone on TiO$_2$ nanostructure at atomistic level. In addition, the most suitable adsorption site of acetone on TiO$_2$ base material can be concluded only after studying the percentage of average energy gap variation ($E_{\text g}^{\text a}$,~\%) compared with its isolated counterpart. Table~\ref{table 3} refers the Mulliken charge transfer, percentage of average energy gap variation and adsorption energy. From the results, it is clearly observed that the prominent adsorption sites for acetone molecule on TiO$_2$ nanostructure are positions Q, R and S. The average energy gap variation is observed to be comparatively higher than P site. The transfer of electron between acetone molecule and TiO$_2$ nanostructure can be analyzed in terms of Mulliken population analysis (Q) \cite{36,37,38}. 

\begin{table}[!t]
\begin{center}
\caption{Adsorption energy, Mulliken charge and average energy gap variation of TiO$_2$ nanostructure.}
\label{table 3}
\vspace{2ex}
\begin{tabular}{|c|c|c|c|c|}
\hline\hline
Nanostructures&$E_{\text{ad}}$ (eV) &Q (e)&$E_{\text g}$ (eV)&$E_{\text g}^{\text a}$ (\%) \\
\hline\hline
\multicolumn{5}{|c|}{Acetone adsorbed on TiO$_2$ nanostructure} \\
\hline
Isolated TiO$_2$&$-$&$-$&2.43&$-$ \\
\hline
P&$-$1.48&0.012&2.32&4.74 \\
\hline
Q&$-$1.46&0.105&1.32&84.09 \\
\hline
R&$-$1.46&0.171&0.82&196.34 \\
\hline
S&$-$1.29&0.013&0.68&257.35 \\
\hline
\multicolumn{5}{|c|}{H$_2$O and O$_2$ adsorbed on TiO$_2$ nanostructure} \\
\hline
A&$-$1.62&0.224&2.39&1.67 \\
\hline
B&$-$1.60&0.17&2.41&0.83 \\
\hline
C&$-$1.78&0.029&2.18&11.47 \\
\hline
D&$-$1.71&$-$0.304&2.36&2.97 \\
\hline\hline
\end{tabular}
\end{center}
\vspace{-4mm}
\end{table}

\begin{figure}[!t]
\centerline{\includegraphics[width=0.7\textwidth]{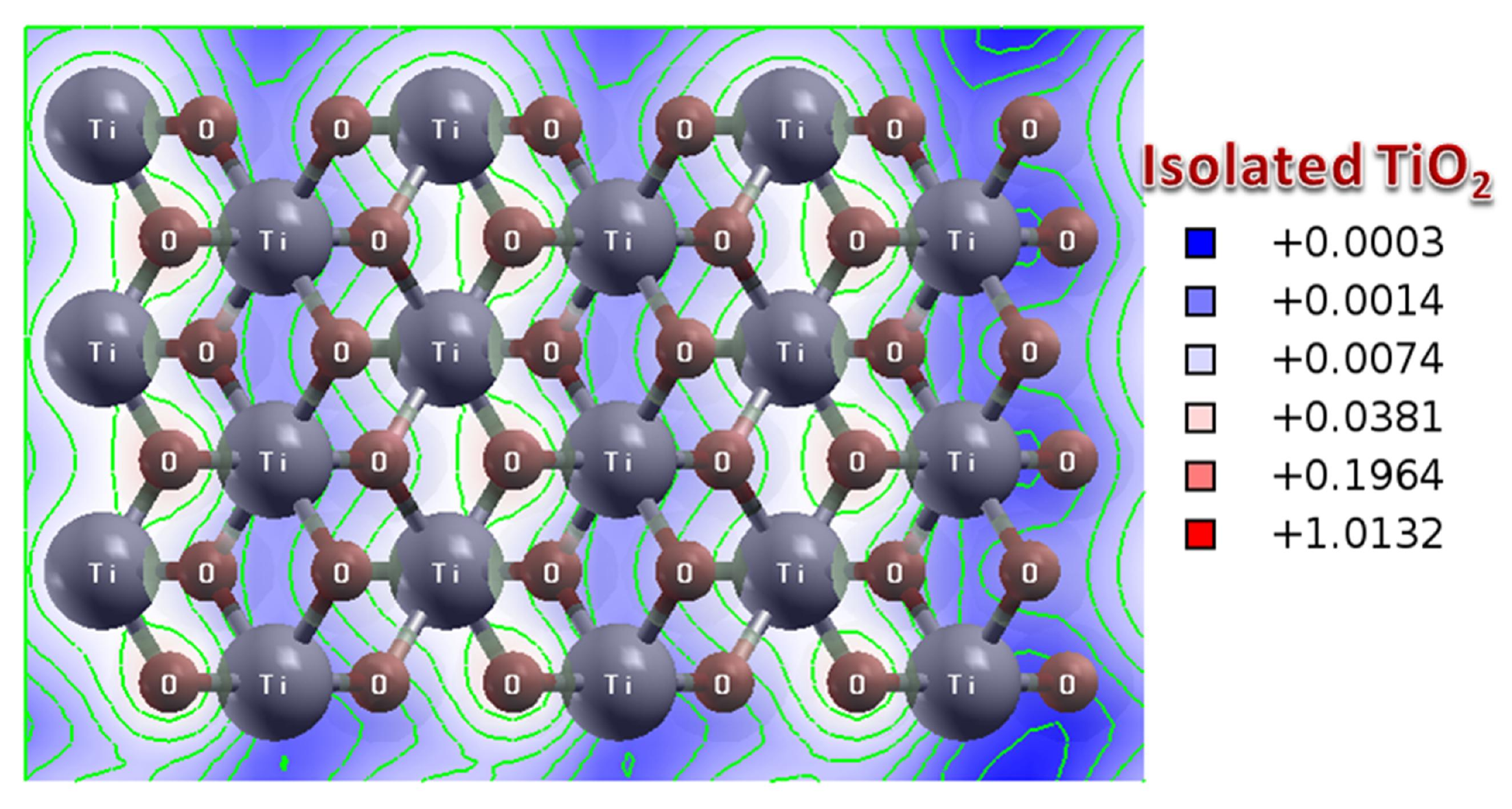}}
\caption{(Color online) Electron density of an isolated TiO$_2$ nanostructure.} \label{fig-s5}
\end{figure}

\begin{figure}[!t]
\centerline{\includegraphics[width=0.99\textwidth]{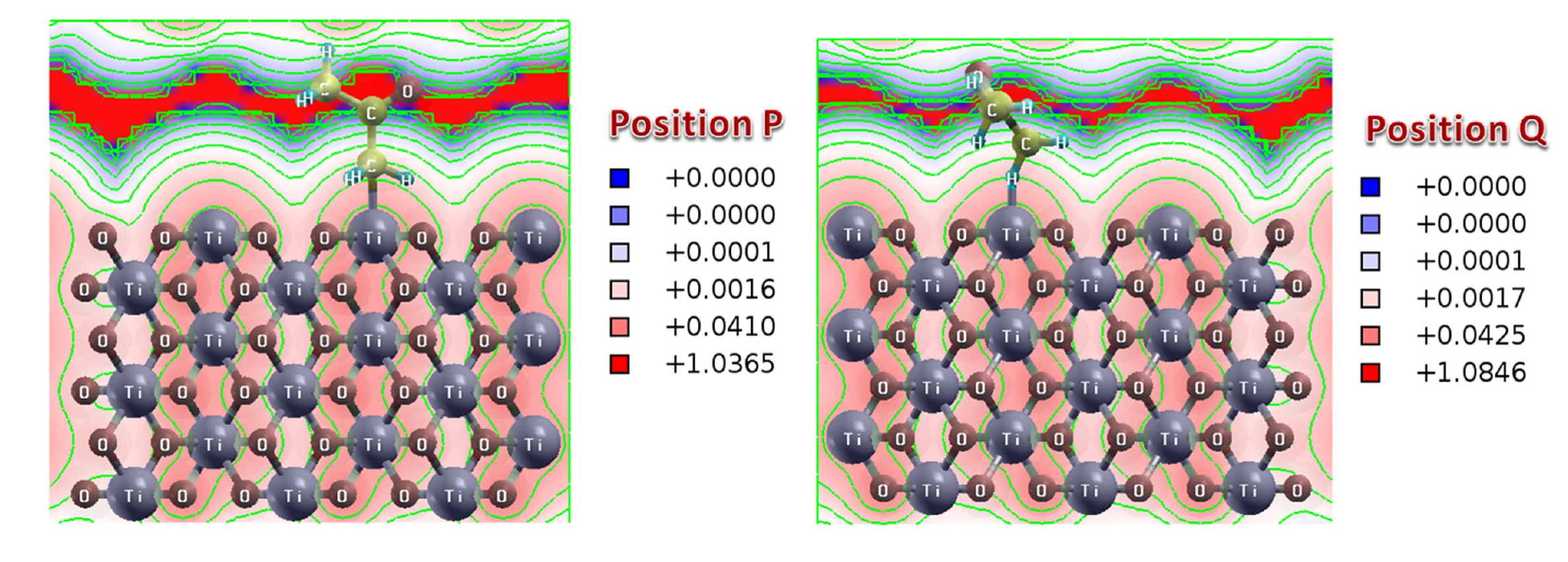}}
\caption{(Color online) Electron density of position P and Q.} \label{fig-s6}
\end{figure}

\begin{figure}[!t]
\centerline{\includegraphics[width=0.99\textwidth]{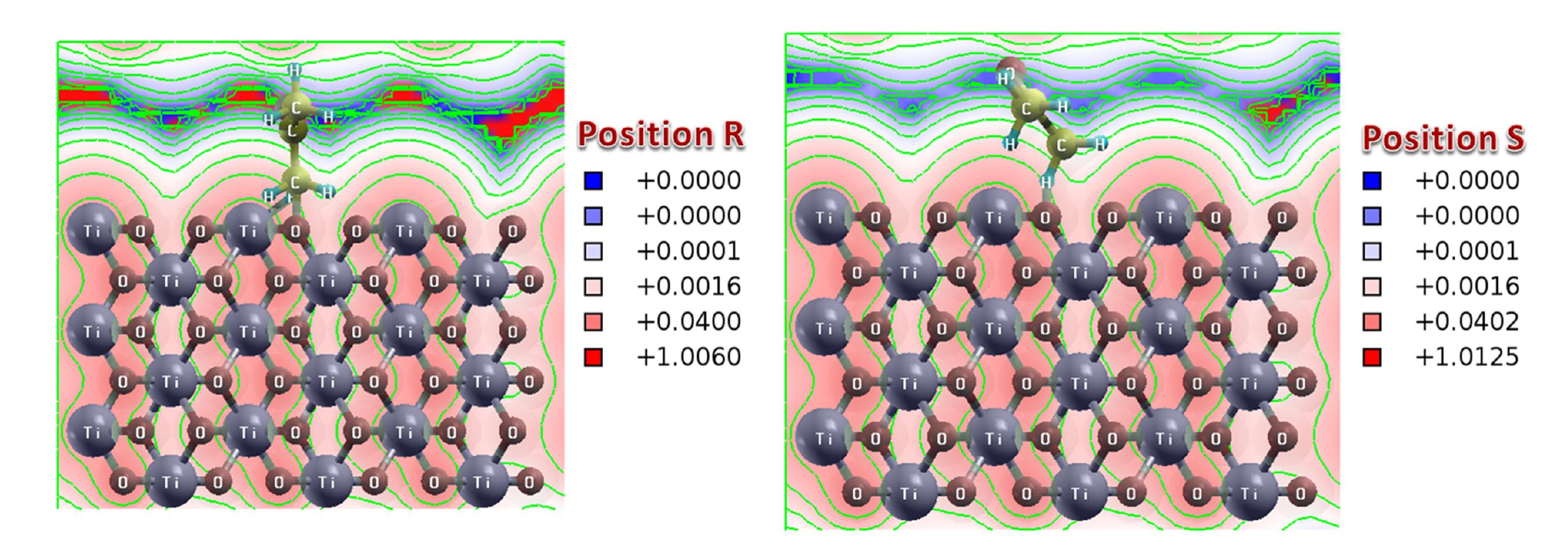}}
\caption{(Color online) Electron density of position R and S.} \label{fig-s7}
\end{figure}

The positive value of Q show that the electrons are transferred from acetone vapor molecule to TiO$_2$ base material; whereas the negative value of Mulliken charge shows that the electrons are transferred from TiO$_2$ base material to acetone molecule \cite{39,40,41}. The Mulliken charge transfer values of TiO$_2$ nanostructure for positions P--S are found to be 0.012~e, 0.105~e, 0.171~e and 0.013~e, respectively. From the observation, the positive value of Mulliken charge is recorded for all the positions upon adsorption of acetone on TiO$_2$ base material. Therefore, the concentration of electrons gets increased due to the transfer of electrons from acetone to TiO$_2$ base material \cite{42,43,44,45}. Moreover, the decrease in the energy gap and transfer of electrons from acetone molecule to TiO$_2$ nanostructures leads to an increase in the current flowing across the two probe TiO$_2$ device. The TiO$_2$ base material can be used for the design and development of acetone sensor. Figures~\ref{fig-s5}--\ref{fig-s7} represents the electron density of an isolated TiO$_2$ nanostructure and for positions P, Q, R and S. 

\begin{figure}[!t]
\centerline{\includegraphics[width=0.99\textwidth]{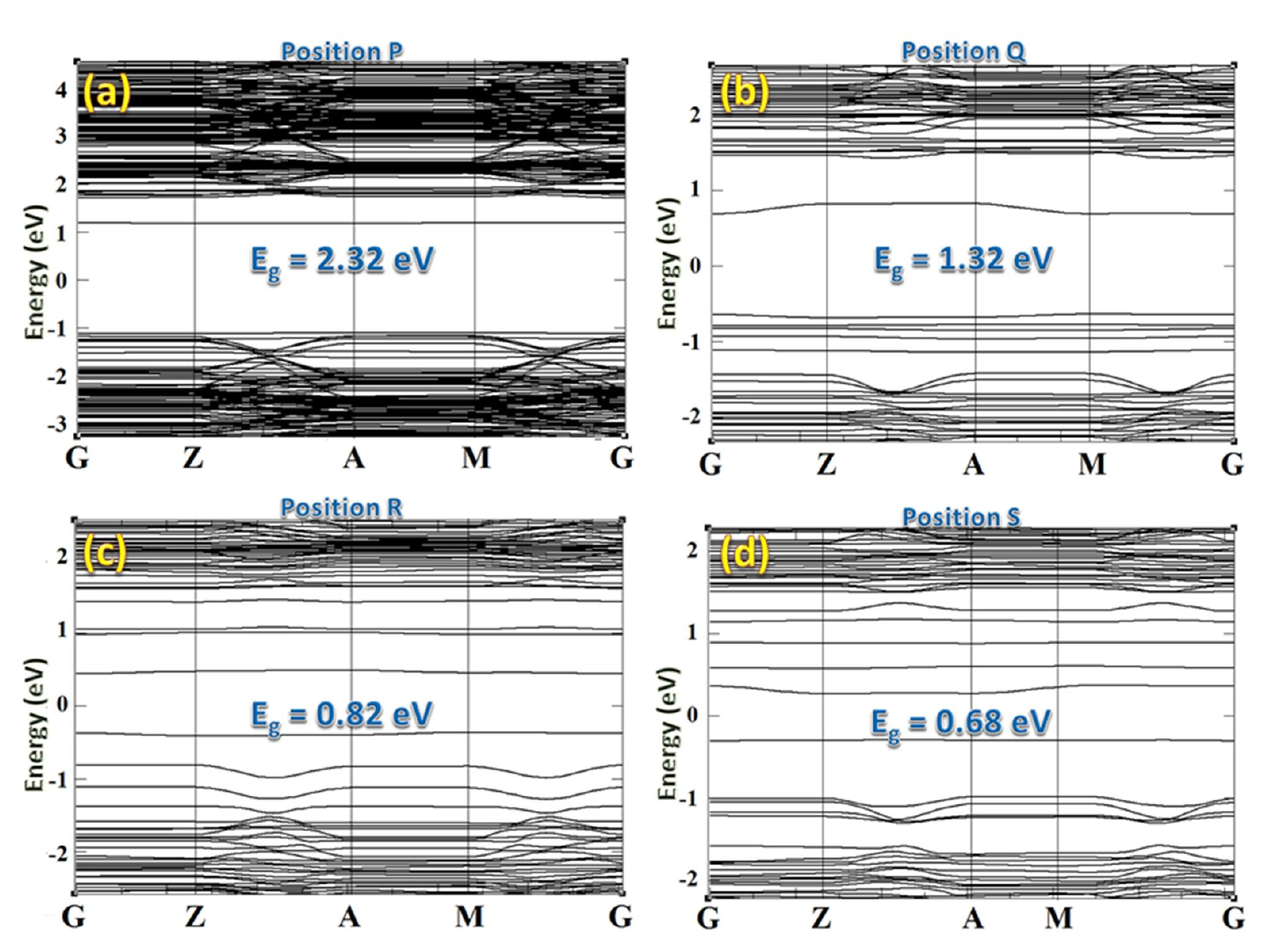}}
\caption{(Color online) (a)--(d) Band structure of TiO$_2$ nanostructure for position P, Q, R and S.} \label{fig-s8}
\end{figure}

\begin{figure}[!t]
\centerline{\includegraphics[width=0.99\textwidth]{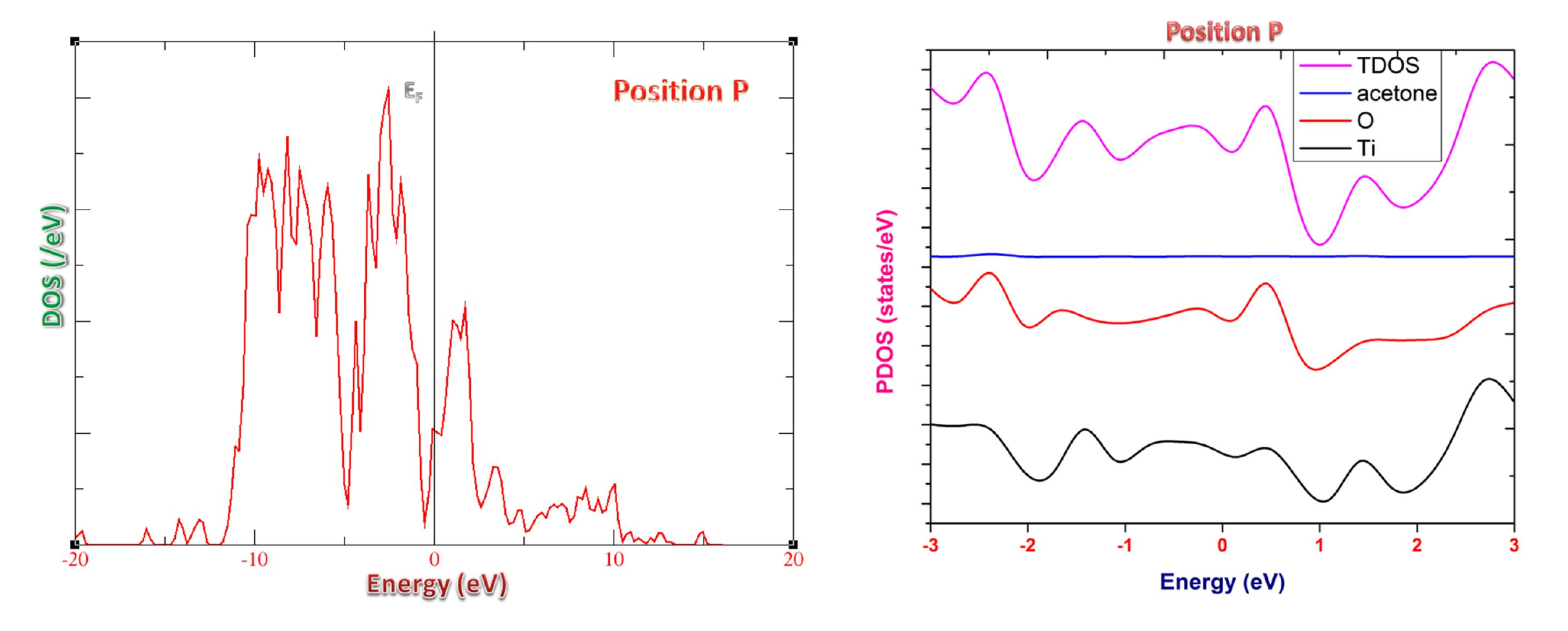}}
\caption{(Color online) Projected density of states (PDOS) and (DOS) spectrum of position P. } \label{fig-s9}
\end{figure}

\begin{figure}[!t]
\vspace{1mm}
\centerline{\includegraphics[width=0.99\textwidth]{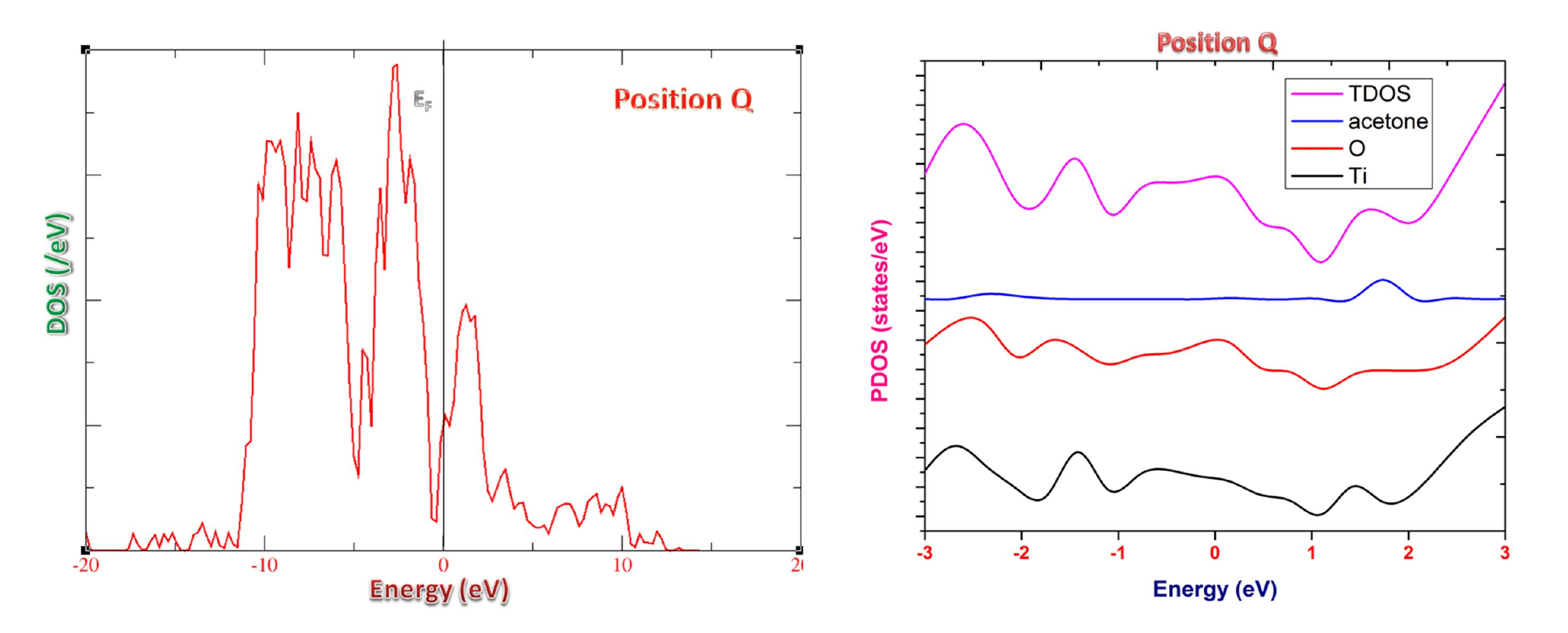}}
\caption{(Color online) Projected density of states (PDOS) and (DOS) spectrum of position Q. } \label{fig-s10}
\end{figure}

\begin{figure}[!t]
\vspace{1mm}
\centerline{\includegraphics[width=0.99\textwidth]{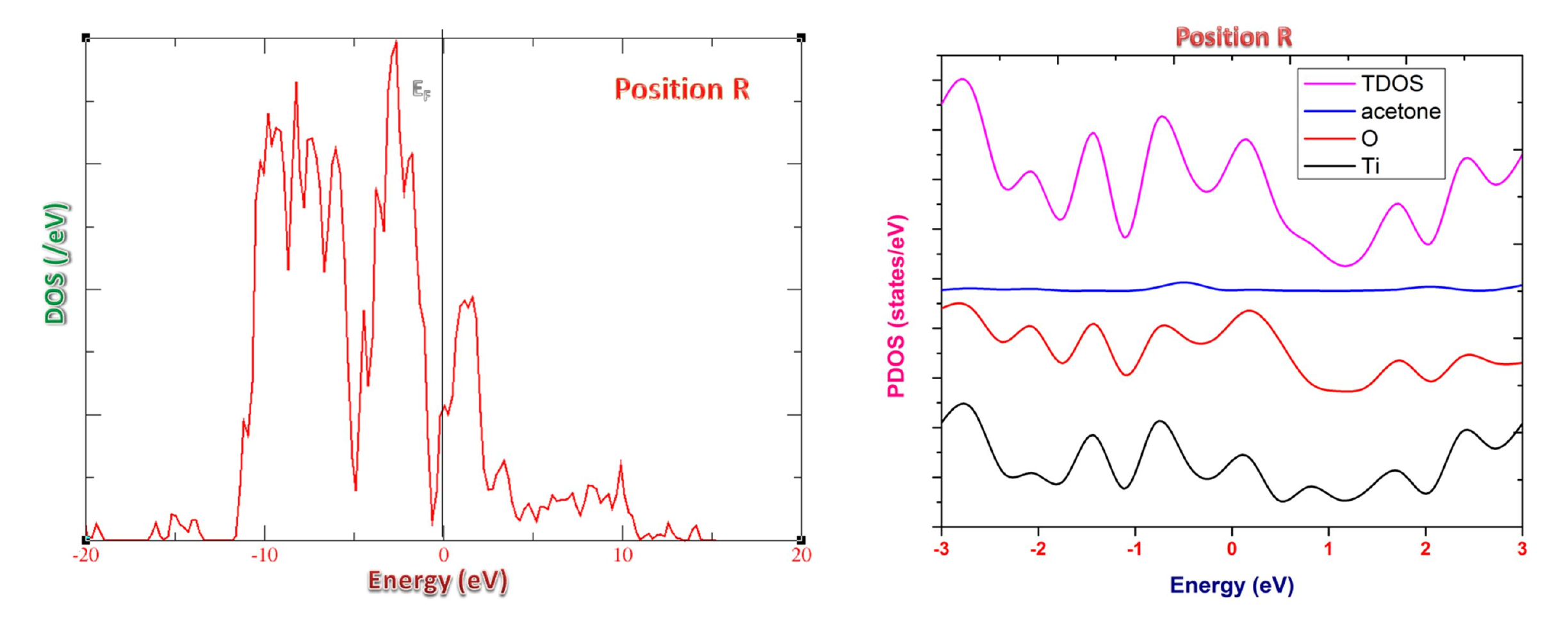}}
\caption{(Color online) Projected density of states (PDOS) and (DOS) spectrum of position R. } \label{fig-s11}
\end{figure}

\begin{figure}[!t]
\centerline{\includegraphics[width=0.99\textwidth]{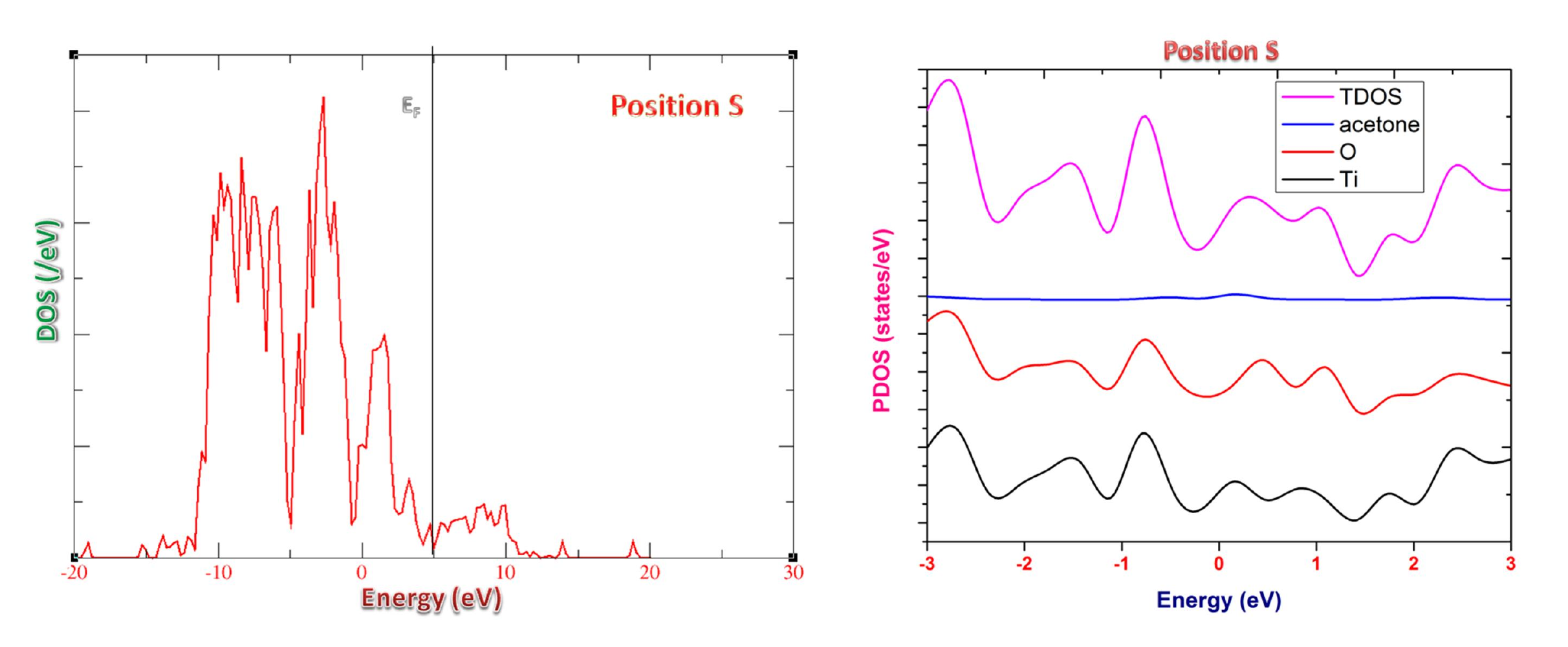}}
\caption{(Color online) Projected density of states (PDOS) and (DOS) spectrum of position S. } \label{fig-s12}
\end{figure}

The variation in the electron density of an isolated TiO$_2$ nanostructure and acetone adsorbed TiO$_2$ nanostructure clearly infers that transition of electrons takes place between acetone and TiO$_2$ nanostructure, which is also in agreement with Mulliken charge transfer. The average energy gap value is found to be high for positions R and S. Also the adsorption energy, Mulliken charge transfer and energy gap variation are found to be favorable for position R. However, for position S, the average energy gap variation is comparatively high with low values of Q and $E_{\text{ad}}$. The Mulliken charge transfer for position P is almost the same as that of position S, whereas the average energy gap variation is comparatively very low. Figure~\ref{fig-s8}~(a)--(d) illustrates the band structure of TiO$_2$ nanostructure for positions P--S, respectively. Figures~\ref{fig-s9}--\ref{fig-s12} represent the corresponding PDOS including DOS spectrum for position P, Q, R and S. From the observation of energy band diagram for positions P and Q, the energy gap is observed to be around 2.32~eV and 1.32~eV, respectively, along the gamma point (G). In the case of position R and S, the energy gap is observed to be around 0.82~eV and 0.68~eV, respectively. Comparing the changes in the energy gap upon adsorption of acetone on TiO$_2$ nanostructure, the variation is observed to be significant for positions Q, R and S. Meanwhile, on looking at the density of states (DOS) spectrum for positions Q, R and S, the density of charge is observed to be larger than an isolated TiO$_2$ counterpart. (DOS spectrum is drawn in multi-curve fashion, the magnitude is taken into consideration along $y$-axis). 

An increase in the density of charge is due to the fact that since TiO$_2$ is n-type semiconductor, the adsorption of acetone molecules consequently gives rise to a transfer of electrons between the acetone vapor and TiO$_2$ base material. The transfer of electrons will increase the electron concentration in TiO$_2$ base material, subsequently increasing the density of charge, which is observed in DOS spectrum. The density of states spectrum gives the insight that the electrons can freely transfer between acetone molecule and TiO$_2$ nanostructure, which can be used as chemical sensor. Thus, the DOS spectrum of TiO$_2$ nanostructure strongly supports the adsorption of acetone molecule on TiO$_2$ material. In order to strengthen the above result, the partial (or projected) density of states spectrum (PDOS) of an isolated TiO$_2$ nanostructure with adsorption sites for the positions P--S is illustrated in figures~\ref{fig-s9}--\ref{fig-s12}. Moreover, the density of charges is noticed to be larger in positions Q, R and S  compared to an isolated TiO$_2$ nanostructure.

\begin{figure}[!b]
\centerline{\includegraphics[width=0.75\textwidth]{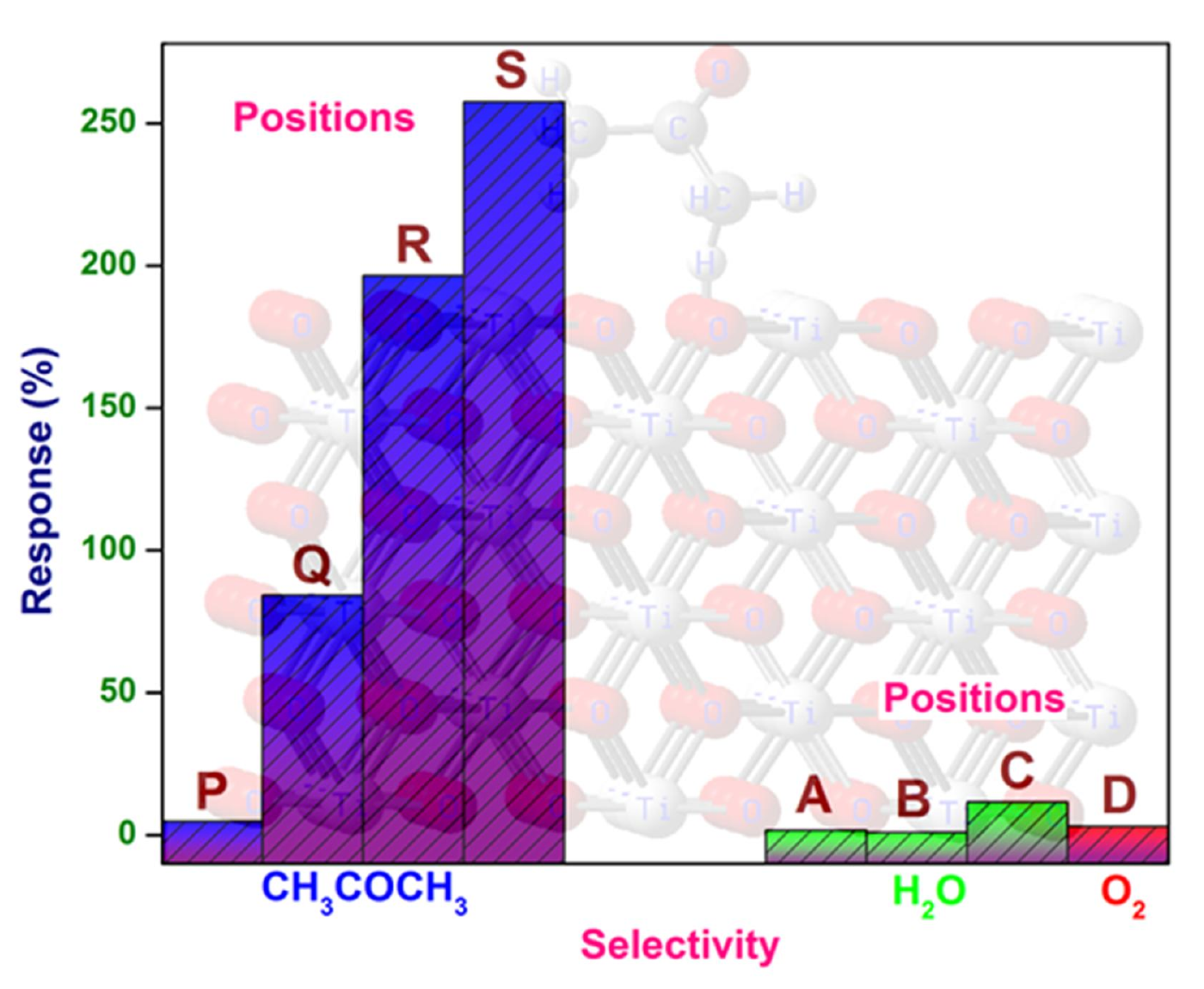}}
\caption{(Color online) The coss-selectivity of acetone in ambient condition with H$_2$O (humidity) and O$_2$ gas molecules. } \label{fig-s13}
\end{figure}

Furthermore, a prominent adsorption site of acetone vapor on TiO$_2$ nanostructure can be concluded only after analyzing the adsorption energy, Mulliken charge transfer, average energy gap variation and energy band gap. From the observations, among all the positions it is found that when hydrogen atom in acetone gets adsorbed on O atom in TiO$_2$ nanostructure, position S is found to be more prominent. Moreover, the selectivity of TiO$_2$ nanostructure towards acetone in the presence of other interfering gases in the ambient condition is to be ascertained. It is well known to the sensor community that sensitivity, selectivity and stability are  important parameters to decide the performance of metal oxide based gas sensors.  Figure~\ref{fig-s13} depicts the selectivity of acetone in ambient condition with other interfering gases, namely H$_2$O (humidity) and O$_2$ gas molecules. The adsorption studies of O$_2$ and H$_2$O on TiO$_2$ nanostructure are carried out and the response towards O$_2$ and H$_2$O molecules is observed.

\begin{figure}[!t]
\centerline{\includegraphics[width=0.9\textwidth]{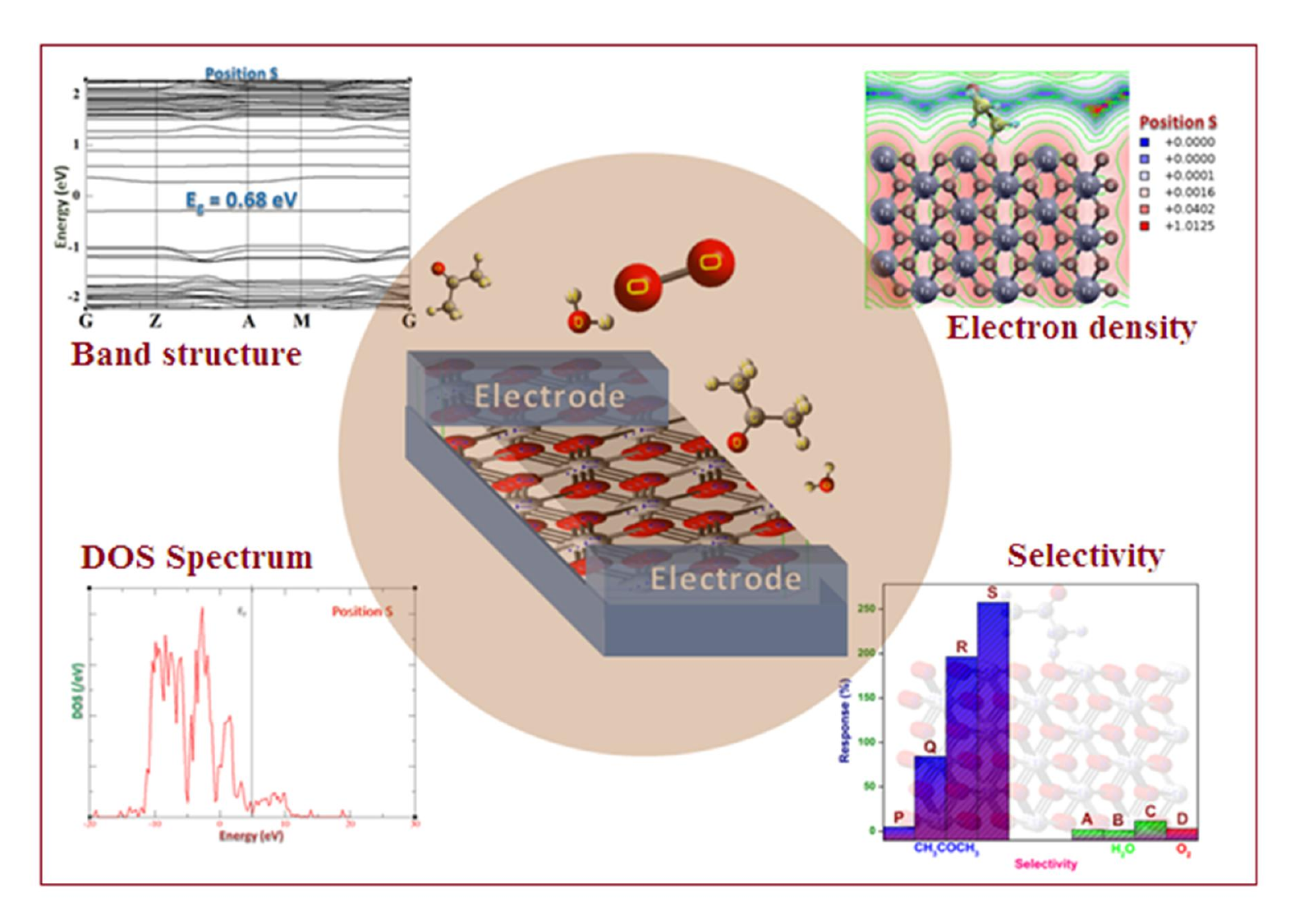}}
\caption{(Color online) The insight on the adsorption behavior of acetone on TiO$_2$ nanostructure. } \label{fig-s14}
\end{figure}

The results show that the response of TiO$_2$ nanostructure towards acetone is found to be relatively high compared with H$_2$O and O$_2$ molecules. Besides, band gap variation upon adsorption of H$_2$O and O$_2$ molecules on TiO$_2$ nanostructure is found to be low. Figure~\ref{fig-s14} shows the insight on the adsorption behavior of acetone on TiO$_2$ nanostructure, which can be used to design a simple two probe device for a possible detection of acetone vapors present in the atmosphere.

\section {Conclusions}

To sum up, the electronic properties and acetone adsorption properties on TiO$_2$ nanostructure are studied using density functional theory method, which is carried out with GGA/PBE exchange-correlation functional.  The band gap of an isolated TiO$_2$ nanostructure is found to be around 2.43~eV. Furthermore, the adsorption of acetone molecules on TiO$_2$ nanostructures is confirmed by the change in the adsorption energy, Mulliken charge transfer and average energy gap variation. The results of DOS spectrum clearly reveal that the transfer of electrons takes place between acetone vapor and TiO$_2$ base material. Furthermore, a prominent adsorption site is explored at atomistic level, which confirms that when a hydrogen atom in acetone molecule gets adsorbed on the O atom in TiO$_2$ nanostructure, the adsorption is found to be favorable. Furthermore, the selectivity of TiO$_2$ nanostructure towards acetone molecules with other interfering gases, namely H$_2$O and O$_2$, is also studied and reported.  The findings from the present investigation strongly support that TiO$_2$ nanostructure can be efficiently used to detect the presence of acetone vapor in the mixed environment. Thus, we conclude that a properly tailored TiO$_2$ nanostructure can be used as a two-probe device to detect the presence of acetone vapors.

\newpage
\ukrainianpart
\title{Детектор для визначення поведінки випарів ацетону на  TiO$_2$ наноструктурах --- застосування теорії функціоналу густини }%
\author{В. Нагараджан, С. Срірам, Р. Чандірамулі }
\address{
Школа електротехніки та електроніки, Академія мистецтв, наукових і технологічних досліджень  Шанмуга (університет SASTRA), Танджавур, Таміл-Наду --- 613 401, Індія}

\makeukrtitle

\begin{abstract}
Досліджуються електронні властивості TiO$_2$  наноструктури з використанням теорії функціоналу густини. Властивості адсорбції  ацетону на  TiO$_2$ 
наноструктурі вивчаються в термінах енергії адсорбції,  зміни середньої енергії щілини і переносу заряду Муллікена. 
Спектр густини станів та структура зони чітко вказують на адсорбцію ацетону на  TiO$_2$  наноструктурах. 
Зміна енергії щілини та зміни густини заряду спостерігаються після адсорбції ацетону на базовому матеріалі   TiO$_2$ n-типу. 
Результати спектру густини станів  показують, що перенос електронів відбувається між парою  ацетону і  TiO$_2$ базовим матеріалом.  
Отримані дані показують, що властивості адсорбції ацетону є більш сприятливими на  TiO$_2$ наноструктурі. 
Зручні місця для адсорбції ацетону на  TiO$_2$ наноструктурі ідентифікуються на атомарному рівні.
 Згідно з отриманими результатами, підтверджується, що  TiO$_2$ 
 наноструктуру можна ефективно використовувати в якості чутливого елемента для виявлення випарів ацетону у змішаному довкіллі.
\keywords TiO$_2$, наноструктура, адсорбція, ацетон, енергетична щілина %
\end{abstract}
\end{document}